\documentclass[10pt,twocolumn]{article}
\usepackage{amsmath,amssymb,amsthm}
\usepackage{mathtools}
\usepackage[dvipsnames]{xcolor}
\usepackage{booktabs}
\usepackage{geometry}
\usepackage{parskip}
\usepackage{titlesec}
\usepackage[numbers,sort&compress]{natbib}
\usepackage{times}
\usepackage{hyperref}

\hypersetup{colorlinks=true, linkcolor=blue!60!black,
            citecolor=blue!60!black, urlcolor=blue!60!black}

\titleformat{\section}[runin]{\bfseries}{}{0pt}{}[.\quad]
\titlespacing{\section}{0pt}{6pt}{0pt}

\begin{document}

\twocolumn[{%
\begin{@twocolumnfalse}
\begin{center}
  {\Large\bfseries 
Schwarzschild solution from a Raychaudhuri-Huygens relation 
  }\\[1em]
  {\normalsize Maurice H.P.M.\ van Putten}\\[0.3em]
  {\small 
  %Department of Physics and Astronomy, Sejong University, Seoul, Republic of Korea
  %\\
  INAF-Osservatorio Astronomico di Capodimonte, Salita Moiariello 16, I-80131 Napoli, Italy\\
  \texttt{mvp@sejong.ac.kr}}\\[1em]
\end{center}

\begin{abstract}
\noindent
Null-congruences encode the causal structure of spacetime, describing a phase space of radiation channels, whose expansion scalar $\theta$ satisfies the Raychaudhuri equation.
In spherical symmetry, the expansion scalar of wave fronts of area $A=4\pi r^2$ satisfies $\theta = A^\prime/A = 2/r$.
About a mass $M$, this carries an encoding area $A_E = \lambda \varphi \ell_p^2 = \lambda R_g r \le A$ in a UV-IR consistent coupling $\ell_p^2= G\hbar/c^3$ to its Compton phase $\varphi = (mc/\hbar)r \propto 1/\hbar$. 
Here, $\lambda = 8\pi $ is fixed by the Newtonian limit at large distances and the saturation limit $A_E=A$ at horizon about $M$ based on the non-local extension of the equivalence principle in Rindler spacetime, without using the Einstein equations. 
In spherical symmetry, this recovers the Schwarzschild metric from a geometric embedding of $p=A_E/A$, which is directly observable in gravitational lensing.
The result shows a non-thermal origin of the scaling dimension of two of gravitational phase space, previously derived based on the second law of thermodynamics by Bekenstein (1973), with possible implications for cosmological spacetime.
\end{abstract}

\hrule\vspace{1em}
\end{@twocolumnfalse}
}]

\noindent\textbf{1.\ Introduction}\quad

Gravitational collapse is a central theme
in modern astronomy. Following core-collapse, 
the end of stellar evolution is believed to 
produce a compact remnant. Massive stars in 
particular may produce a supernova 
\citep{sal55,Lipunov2007,Tonry2018,Graham2019,Jones2021,Aleo2023} leaving behind a neutron star 
or stellar mass black hole \citep{Smartt2009,Sukhb2016,Park2022}. 
Today, we observe black holes in numerous high-energy sources, in binary black hole mergers \citep{ligo2025} and as supermassive objects in the center of galaxies such as M87 \citep[][]{cui23} including the Milky Way \citep[][]{yus26}.

While the existence of astrophysical black holes 
is observationally well established, 
their evolution and physical nature remain a frontier topic as central engines of multi-messenger transient sources \citep[e.g.][]{mvp2024} and theoretical implications for quantum gravity.

As non-trivial solutions to vacuum spacetime with  singularities hidden from view \citep{pen65}, black holes as objects in Nature raise questions on the underlying phase space of spacetime, parameterized in general relativity by Newton's constant $G$ and the velocity of light $c$, and a UV-IR-consistent coupling of matter by Compton phase \citep{van24Ch}, 
parameterized by the reduced Planck constant $\hbar$ and $c$. 

Here, we consider the phase space of null-congruences connecting classical spacetime and matter parameterized by $(G,c)$ and, respectively, $(\hbar,c)$.
These null-congruences encode a causal structure of spacetime, supporting the radiation channels observed classically in the propagation of wave fronts subject to the presence of matter. 
This causal structure introduces a concise definition of phase space allowing for a UV-IR-consistent coupling of matter fields with classical spacetime with no appeal appeal to the second law of thermodynamics \citep{jac95,ver11,van12b,dor26} 
following the seminal work of \citep{bek73}.

In spherical geometry, wave motion is expressed by the expansion of the wave front area $A=4\pi r^2$ with distance $r$. The associated expansion scalar $\theta$ satisfies the Raychaudhuri equation \citep{ray55} that, in spherical symmetry, reduces to the area relation
\begin{eqnarray}
    \theta = A^\prime /A = \frac{2}{r}.
\label{EQN_01}
\end{eqnarray}
About a mass $m$, 
the temporal evolution of (\ref{EQN_01}) is exposed to a Compton wave number $mc/\hbar\propto 1/\hbar$. 
UV-IR consistent coupling to classical spacetime requires trading $1/\hbar$ for $G$. This is realized
by the Planck area $\ell_p^2 = G\hbar/c^3$, in the IR product $\varphi \ell_p^2$ or, equivalently, in the UV constraint $A/\ell_p^2$ on the causal phase space within these wave fronts (\ref{EQN_01}).

Specifically, this trade is encoded in an Einstein area
\begin{eqnarray}
    A_E = \lambda \varphi \ell_p^2 = \lambda R_g r \le A,
    \label{EQN_02}
\end{eqnarray}
where $R_g=Gm/c^2$ denotes the gravitational radius of $m$ and $\lambda$ is a normalization constant. 
The ratio $p=A_E/A$ in (\ref{EQN_02}) represents a fraction of phase space
\begin{eqnarray}
 p = \hat{\lambda} \theta R_g
 \label{EQN_03}
\end{eqnarray}
coupled to $m$ satisfying $0\le p \le 1$,
where $\hat{\lambda}=\lambda/8\pi$. 
We express the expansion rate accordingly,
\begin{eqnarray}
 K\equiv \tau \dot{A}/A = \,p\beta,
 \label{EQN_04}
\end{eqnarray}
where $\tau = \hat{\lambda} R_g/c$ 
and $\beta = \dot{r}/c$ $\left(0\le \beta \le 1\right)$.
This formulates wave motion about $m$ in terms of $K$ expressed by $(p,\beta)$.

In previous work \citep{van24c}, we considered horizon formation in gravitational collapse, based on the equivalence principle and continuation of null-geodesics in pre-collapse Minkowski spacetime. Based on an analytic double covering of the observer's Rindler spacetime, this procedure identifies a horizon radius $R_H=2\hat{\lambda}R_g$ without using the Einstein equations (Fig. 1). 
With $p=1$ in (\ref{EQN_03}-\ref{EQN_04}),
$\tau = \hat{\lambda}R_g/c$ in (\ref{EQN_04}), and hence 
\begin{eqnarray}
p=\frac{R_H}{r}.
\label{EQN_p}
\end{eqnarray}
%in (\ref{EQN_03}). 

In \S2, we consider a closure relation for $\beta=\beta(p)$ in the competition for phase space by radiation and coupling to matter.
In spherical symmetry, the net result is seen to be the Schwarzschild line-element when normalized to the Newtonian limit at large distances. We interpret our approach in \S3, reflecting on gravitational lensing and black holes in the Kerr-Schild metric. 
We summarize our findings in \S4.

\begin{figure}
    \centering
    \includegraphics[width=0.85\linewidth]{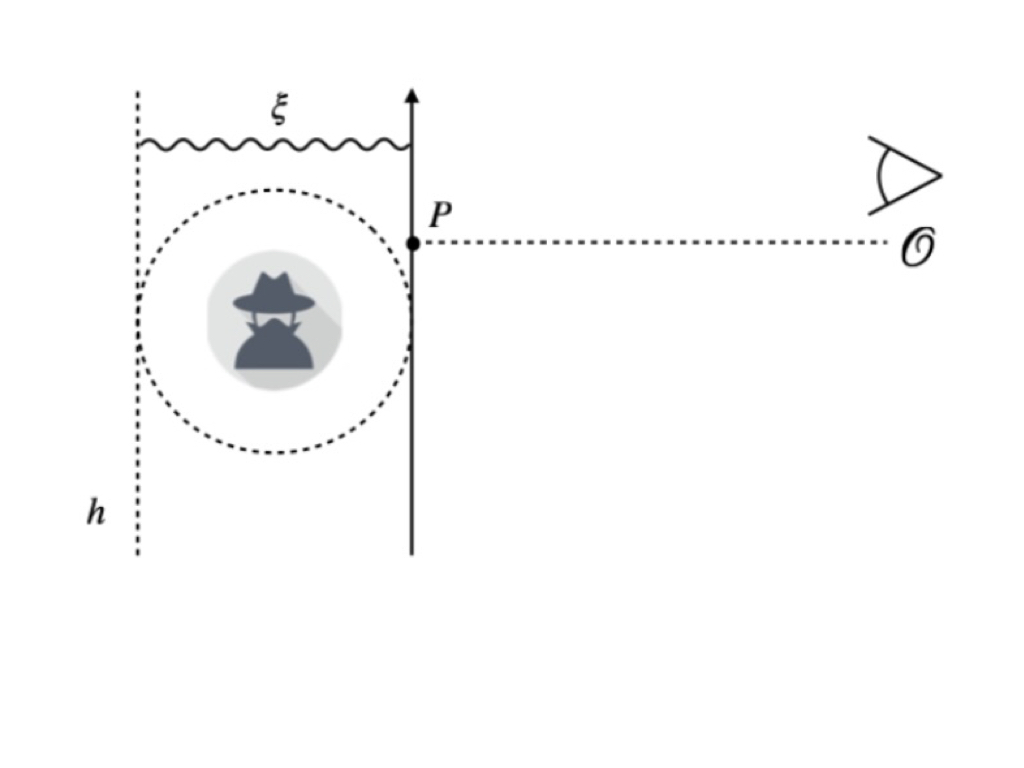}
    \vskip-0.5in
    \caption{
    A distant observer ${\cal O}$ uses a suspended probe $P$ to probe a gravitational field $g=-a$ at acceleration $a$. In ${\cal O}$'s frame of reference, $P$ has an associated Rindler horizon $h$ at a distance $\xi = c^2/a = r^2/(\hat{\lambda}R_g)\ge 2r$.
    In spherical symmetry, the limit $\xi = 2r$ identifies a horizon radius $R_H=2\hat{\lambda}R_g$.
    To ${\cal O}$, the tension in suspending $P$ is equivalently attributed to a Newtonian gravitational attraction $g=-GM/r^2$ by a gravitating mass $M$. [Adapated from \citep{van24c}.]
    }
    \label{fig1}
\end{figure}

\medskip\noindent\textbf{2.\ A Raychaudhuri--Huygens relation}\quad

The symmetric appearance of $0\le p\le 1$ and $0\le \beta\le 1$ in the expansion rate $K$ in (\ref{EQN_04}) with limits $(p,\beta)=(1,0), (0,1)$ suggests a {\it Raychaudhuri-Huygens closure relation} 
\begin{eqnarray}
\beta  =  1 - p = 1 - \frac{R_H}{r}
\label{EQN_21}
\end{eqnarray}
- a minimal closure Ansatz consistent with the boundary conditions.
At the saturation limit $p=1$ in (\ref{EQN_03}), 
$K=0$ in (\ref{EQN_04}), leaving the horizon black.
%(Fig. 2).

This closure (\ref{EQN_21}) describes the competition of scalar expansion $\theta$ 
in regressed wave motion when coupled to matter in the same phase space of null-congruences.
Of the total causal phase constrained by $A/\ell_p^2$, 
a fraction $p$ is occupied by the gravitational encoding of the position of $m$ according to its Compton phase across $r$. The remainder (\ref{EQN_21}) is available for wave propagation. This fractional counting is without free parameters.

At the horizon radius $(p,\beta)=(1,0)$, wave front expansion freezes: $K=0$, consistent with the Penrose trapped-surface condition~\cite{pen65}.
With no causal phase space remaining for wave propagation,the surface is left classically black in the geometric optics approximation.

The closure relation (\ref{EQN_21}) for
the evolution $\dot{r}=\beta c$ along null-geodesics $(ds=0)$ is readily embedded in
a spherically symmetric line element,
\begin{equation}
  ds^2 \;=\; -\beta\,c^2\,dt^2 \;+\; \frac{dr^2}{\beta}
  \;+\; r^2\,d\Omega^2.
  \label{eq:schwarz}
\end{equation}
With (\ref{EQN_21}), imposing Newtonian gravitational attraction at large distances fixes the normalization $\hat{\lambda}=1$. 
Accordingly, (\ref{eq:schwarz}) is the Schwarzschild metric with Schwarzschild radius $R_S=2R_g$.

In our derivation of (\ref{eq:schwarz}), we identify the lapse function in \eqref{eq:schwarz} with the open fraction $\beta = 1-p$ of causal phase space, which defines a gravitational redshift factor $\alpha = \sqrt{\beta}$.

\medskip\noindent\textbf{3.\ Interpretation}\quad

In (\ref{EQN_03}), the normalized gravitational coupling to matter  explicitly factors in
\begin{eqnarray}
p=\theta R_g = \theta \varphi^\prime \ell_p^2
\label{EQN_31}
\end{eqnarray}
over phase space -- in the expansion $\theta$ of null-congruences -- and the UV-IR coupling constant $\ell_p^2$. 
This factoring is explicit also 
in the Kerr--Schild metric
\citep{ker65}
\begin{equation}
  g_{\mu\nu} \;=\; \eta_{\mu\nu} \;+\; p\,k_\mu k_\nu,
 % \qquad p = \frac{\RS}{r},
  \label{eq:KS}
\end{equation}
where $k_\mu = (1,\hat{r})$ is a radially outgoing null vector. 
It explicitly shows a geometric split of the metric in Minkowski background spacetime $\eta_{\mu\nu}$ and a deformation
$p\,k_\mu k_\nu$ encoding $m$, scaled by the UV-IR regularized fraction $p$ of null geodesics of causal phase space.
It identifies $p$ with a probability for a radially outgoing null ray to be gravitationally coupled to $m$ according to the weight by which $m$ deforms the background spacetime along that ray. 

Such is observable in the Einstein area (\ref{EQN_02}) in gravitational lensing. This may be illustrated in a maximally symmetric configuration in which the lens equidistant to source and observer. In this event,
\begin{eqnarray}
    A_E = pA = 8\pi R_g r = \pi \sigma^2 
    \label{EQN_32}
\end{eqnarray}
represents the area of a disk within the Einstein ring of radius $\sigma$ (measured in the source plane). 
For $D_{SL}=D_{L}$, $D_{S}=D_{SL}+D_L$,
the Einstein angle satisfies~\cite{sch92,bar01}
\begin{eqnarray}
\theta_E = \sqrt{ \frac{2R_SD_{SL}}{D_SD_L}}\simeq 0.9^{\prime\prime}
\sqrt{\frac{M_{11}}{D_S}},
\end{eqnarray}
where $M=M_{11}10^{11}M_\odot$ and $D_S$ in units of Gpc.  
With $D_S=2r$ and a lens mass $m$, 
\begin{eqnarray}
A_E=\pi \sigma^2 = 8\pi R_g r
\end{eqnarray}
follows from the ring radius $\sigma=2r \theta_E$.
Typically, $p=A_E/A\ll 1$, and 
$p=\Delta \Omega /4\pi \simeq  \theta_E^2$.

Importantly, $A_E$ in (\ref{EQN_32}) enclosed by the Einstein ring signaling a region of image duplicity. When source, lens, and observer are no longer perfectly aligned, the ring resolves into multiple image pairs, one inside and the other outside the Einstein ring satisfying $\theta_1\theta_2=-\theta_E^2$. This duplicity reflects a redundancy in the imaging process and therefore measures the effective information content associated with encoding the lens position through the fraction $p$ of phase space constrained by $A/\ell_p^2$.

Lensing hereby illustrates the partition of causal phase space between wave motion and the encoding of mass distributions in general relativity described by the mixed hyperbolic–elliptic structure of the Einstein equations \citep[e.g.][]{van96}.

\medskip\noindent\textbf{4.\ Conclusions}\quad

We describe gravitational coupling of matter to classical spacetime mediated by a causal phase space of null-congruences. 
This naturally introduces a UV-IR consistent coupling through
Planck area $\ell_p^2$, providing a geometric brigde between Compton phase and spacetime curvature parameterized in $G$ and $c$. 

In turn, the cross-sectional area of the envelope of null congruences acquires a natural discretization in units of $\ell_p^2$, with geometric scaling dimension $\Delta=2$. The channel capacity associated with null congruences between observer and source combines two transverse area measures, and hence carries an effective geometric scaling dimension $2\Delta=4$. 
Each transverse measure has the same scaling $\Delta_E=2$ as the étendue $E=A\Omega$, where $A$ is the observer’s collecting area and $\Omega$ its aperture, defining a conserved null-ray phase-space throughput.

Accordingly, a mass couples to a fraction $p=A_E/A$ of null geodesics, which
induces a symmetric competition 
(\ref{EQN_04}) between wavefront propagation and regression controlled by $\beta$ and, respectively, $p$, described by the Raychaudhuri--Huygens closure relation (\ref{EQN_21}). 
Embedding this IR-closure relation in the spherically symmetric line element recovers the Schwarzschild metric (\ref{eq:schwarz}).

This gravitational coupling mediated by causal phase space is explicitly manifest in the Kerr--Schild metric (\ref{eq:KS}), 
which is directly observable in gravitational lensing. Both converge on the same parameter $p$, supporting the interpretation of null congruences as a fundamental measure of causal phase space.

Based on the implied Schwarzschild metric, we interpret $p$ as the probability that a null ray gravitationally encodes the position of a mass $m$, representing the weight by which flat spacetime is deformed along that ray. 
Gravity then emerges as the geometric expression of a probability measure defined on null congruences with scaling dimension two, previously identified based on the second law of thermodynamics \citep{bek73}.
This scaling property may have novel implications for dynamical dark energy in cosmological spacetime \citep{van25,abc27,van26,van26Mach}.

\section*{Acknowledgements}
This research is supported in part by NRF grant No. RS-2024-00334550.

\bibliography{PLA-bib}{}
\bibliographystyle{aasjournal}

\end{document}